\documentclass[aps,pra,showpacs,preprint]{revtex4-1}
\usepackage{amsmath}
\usepackage{graphicx}
\usepackage{dcolumn}
\usepackage{bm}
\usepackage{textcomp}
\usepackage{natbib}
\usepackage{upgreek}
\usepackage{enumerate}

\begin{document}
\title{Finding the number density of atomic vapor by studying its absorption profile}
	
\author{Harish Ravi}
\author{Mangesh Bhattarai}
\author{Vasant Natarajan}
\email{vasant@physics.iisc.ernet.in}
	
\affiliation{Department of Physics, Indian Institute of Science, Bangalore-560012, India}

\begin{abstract}
	We demonstrate a technique for obtaining the density of atomic vapor, by doing a fit of the resonant absorption spectrum to a density-matrix model. In order to demonstrate the usefulness of the technique, we apply it to absorption in the ${\rm D_2}$ line of a Cs vapor cell at room temperature. The lineshape of the spectrum is asymmetric due to the role of open transitions. This asymmetry is explained in the model using transit-time relaxation as the atoms traverse the laser beam. We also obtain the latent heat of evaporation by studying the number density as a function of temperature close to room temperature.
\end{abstract}

	\maketitle	

\thispagestyle{empty}
\section{Introduction}
	Theoretical calculations of atomic transition linewidths  \cite{IFN09,CSB11,CHN13,CPB14} and strengths \cite{RTC98,MES99,FOO05} require precise knowledge of the number density of the vapor. While vapor pressure curves are available for most atoms \cite{STE03}, they are not sufficiently accurate to allow them to be used in such calculations. In addition, simplifying assumptions such as the vapor behaving as an ideal gas have to be made in order to relate the vapor pressure to a number density.
	
	In this work, we present an experimental technique to make an accurate estimate of the number density. The basic idea is to use the percentage absorption through a vapor cell. As an example, we demonstrate the usefulness of the technique by measuring the number density of Cs atoms at room temperature. While number density alone is enough to explain the height of the absorption curve, its asymmetric lineshape requires the use of transit-time relaxation as the atoms traverse the laser beam. We also obtain the latent heat of evaporation by studying the number density as a function of temperature.
	
	\section{Experimental details}
\begin{figure}[!h]
			\centering
			\includegraphics[width=0.9\columnwidth]{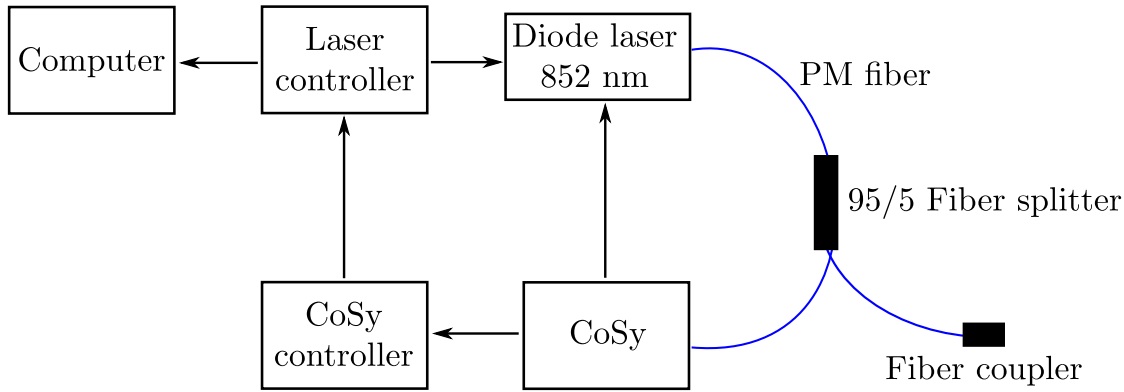}
			\caption{Schematic of the experiment.}
			\label{fig:expsetup}
			\end{figure}

	The experimental set up is shown schematically in Fig.\ \ref{fig:expsetup}. The laser beam is derived from a commercial diode laser system (Toptica DL Pro) operating near the 852 nm $\rm D_2 $ line of ${\rm^{133}Cs}$. The laser output comes out of a single-mode polarization-maintaining (PM) fiber. 
	The fiber goes into a 95/5 power splitter, with $ 5\% $ of the power fed to a Toptica compact spectroscopy (CoSy) unit. The laser controller controls the injection current into the diode and its temperature. Optical feedback to reduce the linewidth of the laser is provided by a piezo-mounted grating, and the piezo voltage is also set by the controller. The laser frequency is scanned by varying the piezo voltage. The controller also acquires data from the CoSy system, which provides a saturated-absorption signal, both with and without a Doppler background. The data is acquired on a computer through a RS232 serial communication link.

	\section{Theoretical analysis}
	\subsection{Density-matrix analysis}
\begin{figure}[!h]
	\centering
	\includegraphics[width=0.5\textwidth]{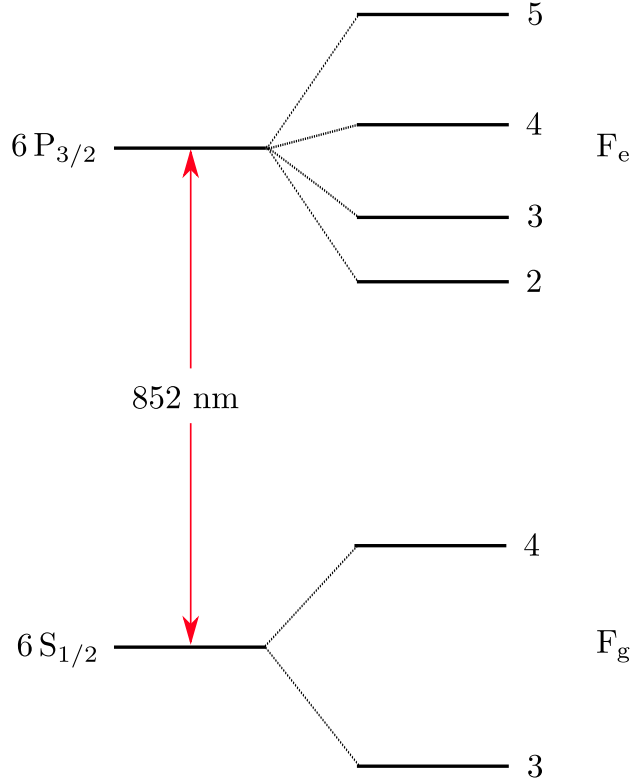}
	\caption{ Hyperfine levels in the $ \rm{D_2} $ line ($ \rm {6\,S_{1/2} \rightarrow 6\,P_{3/2}} $ transition) of $\rm{^{133}Cs} $ (not to scale).}
	\label{levels_Cs}
	\end{figure}

	The absorption of an ideal two-level system is a standard problem discussed in many textbooks \cite{ALE75}. However, transitions in a real atom involve multiple hyperfine levels, which necessitates the use of a numerical density-matrix approach. Closed transitions---those that decay back to the same ground level---behave like two-level systems. On the other hand, open transitions can decay to a different ground level; this necessitates the use of a two-region model---one region where the laser beam is present and one where it is not---something which can be easily incorporated into any numerical package. As an example of the power of this technique, we consider hyperfine transitions in $\rm ^{133}Cs$. The relevant energy levels are shown in Fig.~\ref{levels_Cs}.  Numerical simulations were done using the atomic density-matrix (ADM) package for Mathematica developed by Simon Rochester (can be downloaded from http://rochesterscientific.com/ADM/). It solves numerically the time evolution of the density-matrix elements involved.
	
	We consider transitions starting from the two ground levels of $\rm ^{133}Cs$, namely $ F_g = 3 \rightarrow F_e $ and $ F_g = 4 \rightarrow F_e $, shown schematically in Fig.~\ref{levels_Cs}. In the former case, the $ F_g = 3 \rightarrow F_e = 2 $ transition is closed but it acts like an open system because the atoms get pumped into the extreme magnetic sublevels. For the latter case, only the $ F_g = 4 \rightarrow F_e = 5 $ transitions is closed, while the other two are open. In both cases, modeling of the open transitions requires the use of two regions, and with a relaxation rate to go into the other ground level---transit-time relaxation. The numerical package also allows us to incorporate a rate for atoms to come back to the correct ground level after wall collisions in the region outside the laser beam.
	
	\subsection{Latent heat of evaporation}
	
	The Clausius-Clapeyron equation allows us to extract the latent heat of evaporation of a gas by studying its vapor pressure at two temperatures. It is given by:
	\begin{equation}
	\ln \dfrac{p_1}{p_2} = - \dfrac{L}{R} \left(\dfrac{1}{T_1} - \dfrac{1}{T_2} \right)
	\end{equation}
	where $ p $ is the vapor pressure at temperature $ T $, $ L $ is the latent heat of evaporation, and $ R $ is the universal gas constant. At one temperature, the above equation simplifies to 
	\begin{equation}
	\ln (p) = - \dfrac{L}{R} \dfrac{1}{T} + C
	\end{equation} 
	where the constant $ C $ subsumes all the parameters related to temperature $ T_2 $. We now use the gas law to relate the pressure to the temperature as follows:
	\begin{equation}
	p = n k_B T
	\end{equation}
	where $ n $ is the number density and $ k_B $ is the Boltzmann constant. From this, we get
	\begin{equation}
	\label{nvsT}
	\ln(nT) = - \dfrac{L}{R} \dfrac{1}{T} + {C^{\prime}}
	\end{equation}
	where $ {C^{\prime}} = C - \ln(k_B) $ is a new constant.
	
	\section{Results and discussion}
	
	\subsection{$ F_g = 3 \rightarrow F_e $ transitions}
	
\begin{figure}[!h]
	\centering
	\includegraphics[width=0.6\columnwidth]{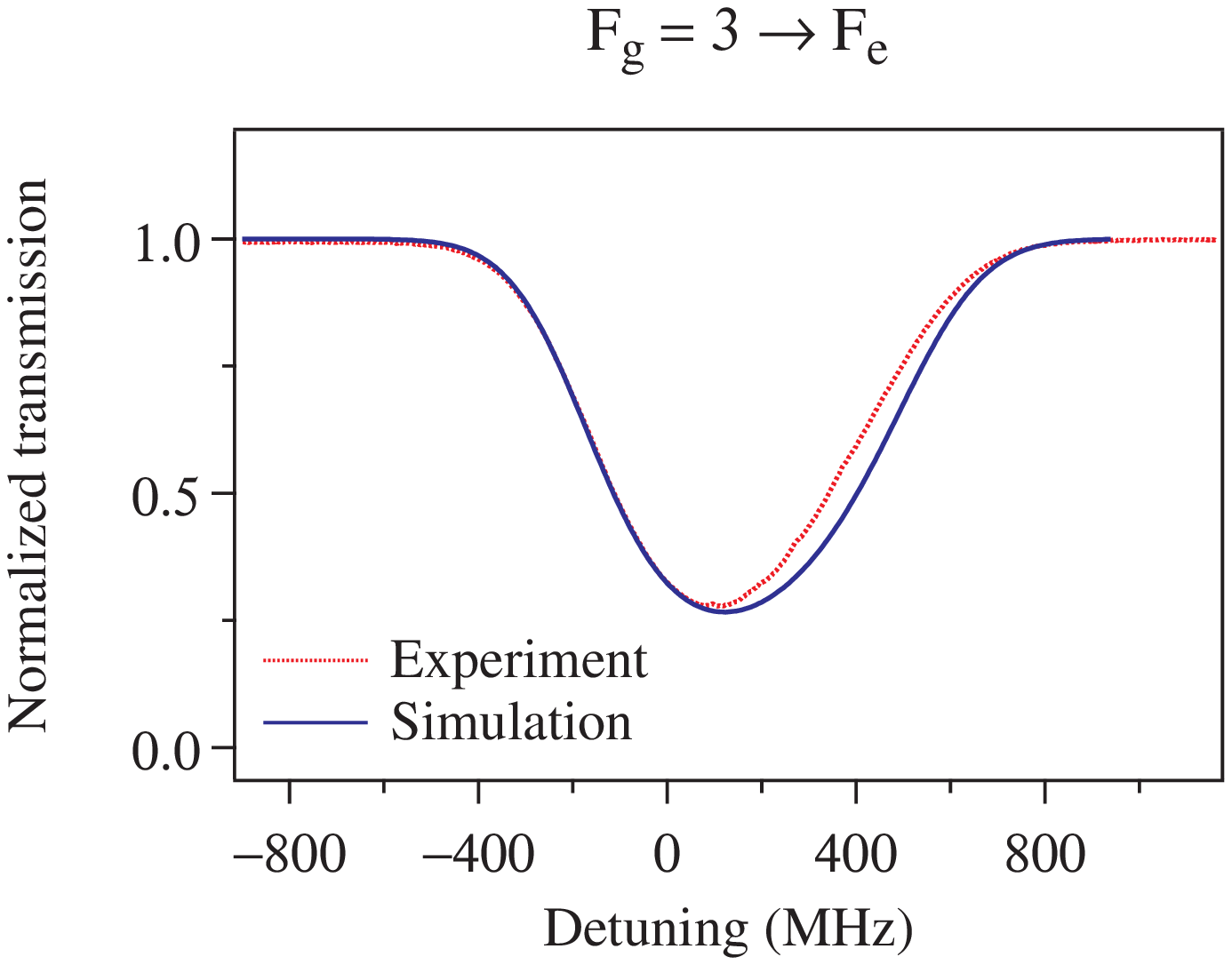}
	\caption{Transmission spectrum of $ 3 \rightarrow F_e $ transitions. The calculated spectrum shows the good match with the experimental one, and is obtai ned with relaxation rates of  $ \gamma_A = 3 \times 10^5 $ s$^{-1}$, $ \gamma_B = 10^2 $ s$^{-1}$, and $ \gamma_w = 10^{10} $ s$^{-1}$.}
	\label{3tox}
	\end{figure}

	We first consider results for transitions starting from the lower ground hyperfine level. Experimental result of a Doppler-broadened spectrum is shown in Fig.~\ref{3tox}. The frequency axis of the laser ($ x $-axis of the figure) is calibrated using separation between hyperfine peaks of a Doppler-corrected saturated-absorption spectrum (not shown). Also shown is the result of the simulation, which shows that it fits the experimental spectrum quite well. The simulation takes as input---(i) the value of $ F_g $, (ii) the light intensity, and (iii) the number density of atoms. It also assumes two regions---A where the laser beam is present, and B where the beam is not present. The number density ($n$) is obtained by varying $n$ manually by trial and error in the simulation, holding all other parameters fixed, until the Doppler broadened curve obtained experimentally matches the theoretical curve.
	
	For open transitions, light in region A pumps atoms into the other $ F_g $ level, while in region B the $ F_g $ levels redistribute due to wall collisions. We assume the relaxation rates to be: $ \gamma_A $ for going from A to B, $ \gamma_B $ for coming back from B to A, and $ \gamma_w $ for redistribution due to wall collisions.
	
	The lineshape of the absorption spectrum  is asymmetric, with the right half behaving like a Gaussian and the left half showing a longer tail. While number density alone is sufficient to give the correct peak height and right half of the curve, relaxation rates $\gamma $'s are required to fit the left half of the curve. This half arises due to open transitions, which is one of the advantages of using a numerical package for the simulation---it allows both the use of a number density and relaxation rates in different regions.
	
\begin{figure}[!h]
		\centering
		\includegraphics[width=0.6\columnwidth]{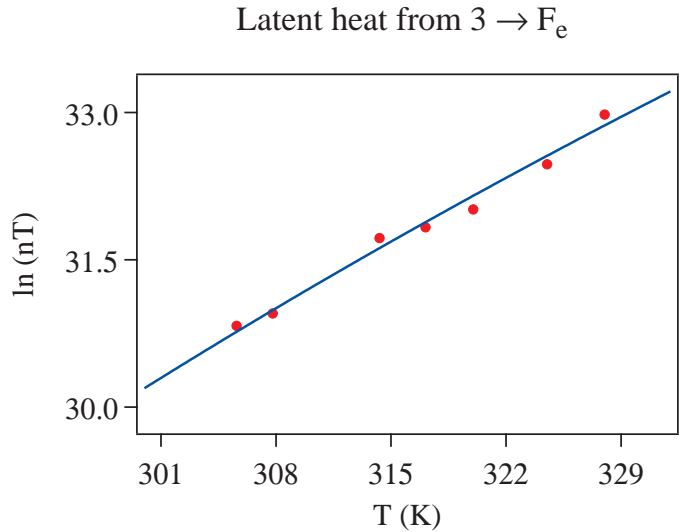}
		\caption{ Variation of number density with temperature for $ 3 \rightarrow F_e $ transitions. The solid line is a fit to Eq.~\eqref{nvsT}, which yields a latent heat of 78(4) kJ/mol.}
		\label{lh3tox}
	\end{figure}

	In the next experiment, the temperature of the CoSy vapor cell was varied and the absorption curve at different temperatures recorded. At each temperature, the curve was fit to a number density. The results of such a measurement are shown in Fig.~\ref{lh3tox}. The symbols are the measured values, while the solid line is a fit to Eq.~\eqref{nvsT}. It is interesting to see that the behavior follows the prediction of Eq.~\eqref{nvsT}. The fit yields a value for the latent heat of evaporation as 78(4) kg/mol. This is a rough value because there are many approximations that go into the deviation of the equation, but the value is comparable to that in other alkali atoms.
	
	\subsection{$ F_g = 4 \rightarrow F_e $ transitions}
	
\begin{figure}[!h]
		\centering
		\includegraphics[width=0.6\columnwidth]{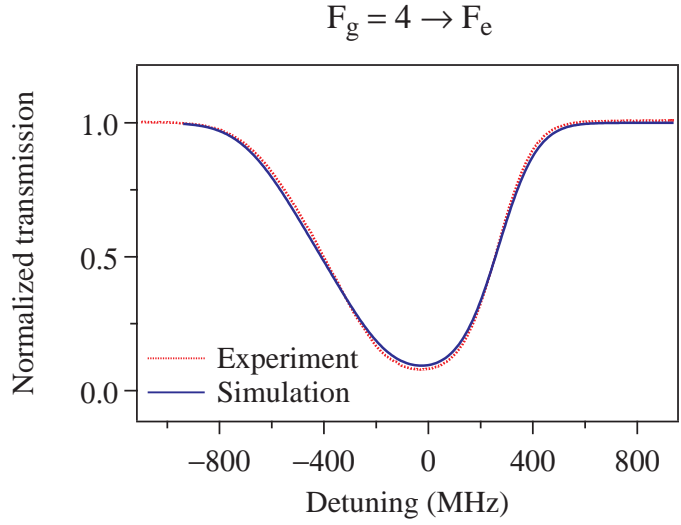}
		\caption{Transmission spectrum of $ 4 \rightarrow F_e $ transitions. The calculated spectrum shows the good match with the experimental one, and is obtained with relaxation rates of  $ \gamma_A = 3 \times 10^5 $ s$^{-1}$, $ \gamma_B = 10^2 $ s$^{-1}$, and $ \gamma_w = 10^{10} $ s$^{-1}$.}
		\label{4tox}
	\end{figure}

	We now consider transitions starting from the upper ground level. We present results of absorption spectra---both experimental and calculated---in Fig.~\ref{4tox}. As in the previous case, experiment and theory match quite well, further validating the numerical package that we use.
	
\begin{figure}[!h]
		\centering
		\includegraphics[width=0.6\columnwidth]{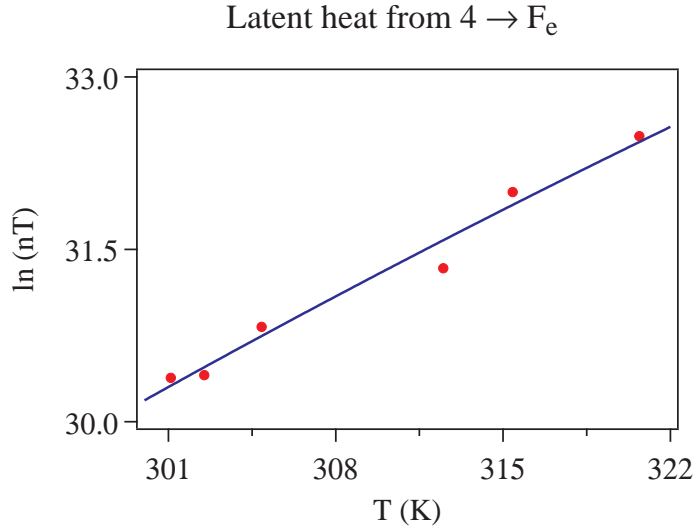}
		\caption{Variation of number density with temperature for $ 4 \rightarrow F_e $ transitions. The solid line is a fit to Eq.~\eqref{nvsT}, which yields a latent heat of 86(7) kJ/mol.}
		\label{lh4tox}
	\end{figure}

	We also present results of variation in number density with temperature, measured in order to get the latent heat of evaporation. The results are presented in Fig.~\ref{lh4tox}. The solid line, which is a fit to Eq.~\eqref{nvsT} yields the latent heat as 86(7) kJ/mol. This value is consistent with the value obtained in the other case.
	
	\section{Conclusions}
	
	In summary, we have demonstrated a technique for getting the number density of atomic vapor, by fitting the resonant absorption spectrum to a numerical density-matrix model. As an example of the power of this technique, we apply it to absorption through a room temperature vapor of Cs atoms. The lineshape of the spectrum is asymmetric due to the role of open transitions. This asymmetry is explained in the model by having two regions---one with and one without the laser beam---along with transit-time relaxation as the atoms traverse the laser beam. We also study the absorption profile as a function of temperature (near room temperature), and combine it with the Clausius-Clapeyron equation, to estimate the latent heat of evaporation.  We demonstrate that our technique works for both hyperfine levels of the ground state of $ \rm {^{133}Cs} $.
	
	\section*{Acknowledgments}
	This work was supported by a grant from the Indian Institute of Science, Bangalore. The authors thank S Raghuveer for help with the manuscript preparation. 


\begin{thebibliography}{9}%
		\makeatletter
		\providecommand \@ifxundefined [1]{%
		 \@ifx{#1\undefined}
		}%
		\providecommand \@ifnum [1]{%
		 \ifnum #1\expandafter \@firstoftwo
		 \else \expandafter \@secondoftwo
		 \fi
		}%
		\providecommand \@ifx [1]{%
		 \ifx #1\expandafter \@firstoftwo
		 \else \expandafter \@secondoftwo
		 \fi
		}%
		\providecommand \natexlab [1]{#1}%
		\providecommand \enquote  [1]{``#1''}%
		\providecommand \bibnamefont  [1]{#1}%
		\providecommand \bibfnamefont [1]{#1}%
		\providecommand \citenamefont [1]{#1}%
		\providecommand \href@noop [0]{\@secondoftwo}%
		\providecommand \href [0]{\begingroup \@sanitize@url \@href}%
		\providecommand \@href[1]{\@@startlink{#1}\@@href}%
		\providecommand \@@href[1]{\endgroup#1\@@endlink}%
		\providecommand \@sanitize@url [0]{\catcode `\\12\catcode `\$12\catcode
		  `\&12\catcode `\#12\catcode `\^12\catcode `\_12\catcode `\%12\relax}%
		\providecommand \@@startlink[1]{}%
		\providecommand \@@endlink[0]{}%
		\providecommand \url  [0]{\begingroup\@sanitize@url \@url }%
		\providecommand \@url [1]{\endgroup\@href {#1}{\urlprefix }}%
		\providecommand \urlprefix  [0]{URL }%
		\providecommand \Eprint [0]{\href }%
		\providecommand \doibase [0]{http://dx.doi.org/}%
		\providecommand \selectlanguage [0]{\@gobble}%
		\providecommand \bibinfo  [0]{\@secondoftwo}%
		\providecommand \bibfield  [0]{\@secondoftwo}%
		\providecommand \translation [1]{[#1]}%
		\providecommand \BibitemOpen [0]{}%
		\providecommand \bibitemStop [0]{}%
		\providecommand \bibitemNoStop [0]{.\EOS\space}%
		\providecommand \EOS [0]{\spacefactor3000\relax}%
		\providecommand \BibitemShut  [1]{\csname bibitem#1\endcsname}%
		\let\auto@bib@innerbib\@empty
		\bibitem [{\citenamefont {Iftiquar}\ and\ \citenamefont
		  {Natarajan}(2009)}]{IFN09}%
		  \BibitemOpen
		  \bibfield  {author} {\bibinfo {author} {\bibfnamefont {S.~M.}\ \bibnamefont
		  {Iftiquar}}\ and\ \bibinfo {author} {\bibfnamefont {V.}~\bibnamefont
		  {Natarajan}},\ }\href {\doibase 10.1103/PhysRevA.79.013808} {\bibfield
		  {journal} {\bibinfo  {journal} {Phys. Rev. A}\ }\textbf {\bibinfo {volume}
		  {79}},\ \bibinfo {pages} {013808} (\bibinfo {year} {2009})}\BibitemShut
		  {NoStop}%
		\bibitem [{\citenamefont {Chanu}\ \emph {et~al.}(2011)\citenamefont {Chanu},
		  \citenamefont {Singh}, \citenamefont {Brun}, \citenamefont {Pandey},\ and\
		  \citenamefont {Natarajan}}]{CSB11}%
		  \BibitemOpen
		  \bibfield  {author} {\bibinfo {author} {\bibfnamefont {S.~R.}\ \bibnamefont
		  {Chanu}}, \bibinfo {author} {\bibfnamefont {A.~K.}\ \bibnamefont {Singh}},
		  \bibinfo {author} {\bibfnamefont {B.}~\bibnamefont {Brun}}, \bibinfo {author}
		  {\bibfnamefont {K.}~\bibnamefont {Pandey}}, \ and\ \bibinfo {author}
		  {\bibfnamefont {V.}~\bibnamefont {Natarajan}},\ }\href {\doibase DOI:
		  10.1016/j.optcom.2011.07.001} {\bibfield  {journal} {\bibinfo  {journal}
		  {Opt. Commun.}\ }\textbf {\bibinfo {volume} {284}},\ \bibinfo {pages} {4957 }
		  (\bibinfo {year} {2011})}\BibitemShut {NoStop}%
		\bibitem [{\citenamefont {Chanu}\ and\ \citenamefont
		  {Natarajan}(2013)}]{CHN13}%
		  \BibitemOpen
		  \bibfield  {author} {\bibinfo {author} {\bibfnamefont {S.~R.}\ \bibnamefont
		  {Chanu}}\ and\ \bibinfo {author} {\bibfnamefont {V.}~\bibnamefont
		  {Natarajan}},\ }\href {\doibase 10.1016/j.optcom.2013.01.042} {\bibfield
		  {journal} {\bibinfo  {journal} {Opt. Commun.}\ }\textbf {\bibinfo {volume}
		  {295}},\ \bibinfo {pages} {150} (\bibinfo {year} {2013})}\BibitemShut
		  {NoStop}%
		\bibitem [{\citenamefont {Chanu}\ \emph {et~al.}(2014)\citenamefont {Chanu},
		  \citenamefont {Pandey}, \citenamefont {Bharti}, \citenamefont {Wasan},\ and\
		  \citenamefont {Natarajan}}]{CPB14}%
		  \BibitemOpen
		  \bibfield  {author} {\bibinfo {author} {\bibfnamefont {S.~R.}\ \bibnamefont
		  {Chanu}}, \bibinfo {author} {\bibfnamefont {K.}~\bibnamefont {Pandey}},
		  \bibinfo {author} {\bibfnamefont {V.}~\bibnamefont {Bharti}}, \bibinfo
		  {author} {\bibfnamefont {A.}~\bibnamefont {Wasan}}, \ and\ \bibinfo {author}
		  {\bibfnamefont {V.}~\bibnamefont {Natarajan}},\ }\href
		  {http://stacks.iop.org/0295-5075/106/i=4/a=43001} {\bibfield  {journal}
		  {\bibinfo  {journal} {Europhys. Lett.}\ }\textbf {\bibinfo {volume} {106}},\
		  \bibinfo {pages} {43001} (\bibinfo {year} {2014})}\BibitemShut {NoStop}%
		\bibitem [{\citenamefont {Rafac}\ and\ \citenamefont {Tanner}(1998)}]{RTC98}%
		  \BibitemOpen
		  \bibfield  {author} {\bibinfo {author} {\bibfnamefont {R.~J.}\ \bibnamefont
		  {Rafac}}\ and\ \bibinfo {author} {\bibfnamefont {C.~E.}\ \bibnamefont
		  {Tanner}},\ }\href {\doibase 10.1103/PhysRevA.58.1087} {\bibfield  {journal}
		  {\bibinfo  {journal} {Phys. Rev. A}\ }\textbf {\bibinfo {volume} {58}},\
		  \bibinfo {pages} {1087} (\bibinfo {year} {1998})}\BibitemShut {NoStop}%
		\bibitem [{\citenamefont {Metcalf}\ and\ \citenamefont {van~der
		  Stratten}(1999)}]{MES99}%
		  \BibitemOpen
		  \bibfield  {author} {\bibinfo {author} {\bibfnamefont {H.~J.}\ \bibnamefont
		  {Metcalf}}\ and\ \bibinfo {author} {\bibfnamefont {P.}~\bibnamefont {van~der
		  Stratten}},\ }\href {http://www.springer.com/gp/book/9780387987286}
		  {\bibfield  {journal} {\bibinfo  {journal} {Springer}\ } (\bibinfo {year}
		  {1999})}\BibitemShut {NoStop}%
		\bibitem [{\citenamefont {Foot}(2005)}]{FOO05}%
		  \BibitemOpen
		  \bibfield  {author} {\bibinfo {author} {\bibfnamefont {C.~J.}\ \bibnamefont
		  {Foot}},\ }\href@noop {} {\bibfield  {journal} {\bibinfo  {journal} {Oxford
		  university press}\ } (\bibinfo {year} {2005})}\BibitemShut {NoStop}%
		\bibitem [{\citenamefont {Steck}(2003)}]{STE03}%
		  \BibitemOpen
		  \bibfield  {author} {\bibinfo {author} {\bibfnamefont {D.~A.}\ \bibnamefont
		  {Steck}},\ } {Cesium D line data, http://steck.us/alkalidata/} {\  (\bibinfo {year}
		  {2003})}\BibitemShut {NoStop}%
		\bibitem [{\citenamefont {Allen}\ and\ \citenamefont {Eberly}(1975)}]{ALE75}%
		  \BibitemOpen
		  \bibfield  {author} {\bibinfo {author} {\bibfnamefont {L.}~\bibnamefont
		  {Allen}}\ and\ \bibinfo {author} {\bibfnamefont {J.~H.}\ \bibnamefont
		  {Eberly}},\ }\href@noop {} {\bibfield  {journal} {\bibinfo  {journal} {Dover
		  publications}\ } (\bibinfo {year} {1975})}\BibitemShut {NoStop}%
		\end{thebibliography}

		%

\end{document}